%% file: main.tex
\documentclass[sigconf]{acmart}

\usepackage{xcolor, graphicx}
\usepackage{color, colortbl}
\usepackage{tcolorbox}
\usepackage[export]{adjustbox}
\usepackage{subcaption}
\usepackage{multirow}
\usepackage{ulem}
\usepackage{enumitem}

\newcommand\BibTeX{B\textsc{ib}\TeX}

\newcommand{\vsum}{MTLTS}
\RequirePackage{xcolor}
\definecolor{imperialblue}{RGB}{0,62,116}
\tcbset{colback=gray!5!white, colframe=imperialblue!75!black, fonttitle=\bfseries, center title}

\AtBeginDocument{%
  \providecommand\BibTeX{{%
    \normalfont B\kern-0.5em{\scshape i\kern-0.25em b}\kern-0.8em\TeX}}}

\copyrightyear{2022}
\acmYear{2022}
\setcopyright{acmcopyright}
\acmConference[WSDM '22] {Proceedings of the Fifteenth ACM International Conference on Web Search and Data Mining}{February 21--25, 2022}{Tempe, AZ, USA.}
\acmBooktitle{Proceedings of the Fifteenth ACM International Conference on Web Search and Data Mining (WSDM '22), February 21--25, 2022, Tempe, AZ, USA}
\acmPrice{15.00}
\acmISBN{978-1-4503-9132-0/22/02}
\acmDOI{10.1145/3488560.3498536}

\acmSubmissionID{wsdmfp870}

\begin{document}
\fancyhead{}

\title{\textit{\vsum}: A Multi-Task Framework To Obtain Trustworthy Summaries From Crisis-Related Microblogs}

\author{Rajdeep Mukherjee}
\affiliation{\institution{IIT Kharagpur, India}}
\email{rajdeep1989@iitkgp.ac.in}

\author{Uppada Vishnu}
\authornote{Both authors contributed equally to this research.}
\affiliation{\institution{IIT Kharagpur, India}}
\email{vishnu14july@iitkgp.ac.in}

\author{Hari Chandana Peruri}
\authornotemark[1]
\affiliation{\institution{IIT Kharagpur, India}}
\email{chandupvsl@iitkgp.ac.in}


\author{Sourangshu Bhattacharya}
\affiliation{\institution{IIT Kharagpur, India}}
\email{sourangshu@cse.iitkgp.ac.in}

\author{Koustav Rudra}
\authornote{Research was conducted while the author was affiliated to L3S Research Center.}
\affiliation{\institution{IIT(ISM) Dhanbad, India}}
\email{koustav@iitism.ac.in}

\author{Pawan Goyal}
\affiliation{\institution{IIT Kharagpur, India}}
\email{pawang@cse.iitkgp.ac.in}

\author{Niloy Ganguly}
\affiliation{\institution{IIT Kharagpur, India \& Leibniz University of Hannover, Germany}}
\email{niloy@cse.iitkgp.ac.in}

\renewcommand{\shortauthors}{Rajdeep Mukherjee, et al.}

\begin{abstract}
Occurrences of catastrophes such as natural or man-made disasters trigger the spread of rumours over social media at a rapid pace. Presenting a \textbf{trustworthy} and \textbf{summarized} account of the unfolding event in near real-time to the consumers of such potentially unreliable information thus becomes an important task. In this work, we propose \textbf{\vsum}, the first end-to-end solution for the task that jointly determines the credibility and summary-worthiness of tweets. Our credibility \textbf{verifier} is designed to recursively learn the structural properties of a Twitter conversation cascade, along with the stances of replies towards the source tweet. We then take a hierarchical multi-task learning approach, where the \textit{verifier} is trained at a lower layer, and the \textbf{summarizer} is trained at a deeper layer where it utilizes the \textit{verifier} predictions to determine the salience of a tweet. Different from existing \textit{disaster-specific} summarizers, we model tweet summarization as a \textit{supervised} task. Such an approach can automatically learn summary-worthy features, and can therefore generalize well across domains. When trained on the \textit{PHEME} dataset \cite{zubiaga2015crowdsourcing}, not only do we outperform the strongest baselines for the auxiliary task of \textit{verification/rumour detection}, we also achieve \textbf{21 - 35\%} gains in the \textit{verified ratio} of summary tweets, and \textbf{16 - 20\%} gains in \textit{ROUGE1-F1} scores over the existing state-of-the-art solutions for the primary task of \textbf{trustworthy summarization}.



\end{abstract}



\maketitle
\input{tables}
\input{intro}
\input{model}
\input{experiments}
\input{results}
\input{analysis}
\input{conclusion}

\begin{acks}
    This research is partially supported by IMPRINT-2, a national initiative of the Ministry of Human Resource Development (MHRD), India. Niloy Ganguly was partially funded by the Federal Ministry of Education and Research (BMBF), Germany (grant no. 01DD20003).
\end{acks}

\newpage
\bibliographystyle{ACM-Reference-Format}
\bibliography{main}

\end{document}

%% file: tables.tex
\newcommand{\tabdataset}{
\begin{table}
    \centering
    \caption{Statistics of the PHEME Dataset.}
    \label{tab:pheme5-satistics}
    \resizebox{\columnwidth}{!}{
    \begin{tabular}{|c|c|c|c|c|}
        \hline
        \textbf{Disaster-Events}  & \textbf{\# Tweets} & \multicolumn{3}{c|}{\textbf{\# of Source Tweets}}  \\
        \cline{3-5}
        & \textbf{w/ Stance} & \textbf{Verified} & \textbf{Unverified} & \textbf{Total} \\
        \hline
        Charlie Hebdo & 1,071 & 1,621 (78.0\%) & 458 (22.0\%) & 2,079  \\
        \hline
        Germanwings-crash & 281 & 231 (49.3\%) & 238 (50.7\%) & 469 \\
        \hline
        Ottawa shooting & 777 & 420 (47.2\%) & 470 (52.8\%) & 890 \\
        \hline
        Sydney seige & 1,107 & 699 (57.2\%) & 522 (42.8\%) & 1,221 \\
        \hline
        \hline
        \textbf{Total} & \textbf{3,236} & \textbf{2,971 (64\%)} & \textbf{1,688 (36\%)} & \textbf{4659} \\
        \hline
    \end{tabular}
    }
    \vspace{-3mm}
\end{table}
}

\newcommand{\tabmtlsummRichness}{
\begin{table*}
    \centering
    \caption{Qualitative comparison of summaries generated by various methods. Overall, \vsum \hspace{1pt} achieves best \textit{Verified Ratio} of summary tweets for $\kappa=0$ and best ROUGE-1 F-1 scores for $\kappa=1$. \vsum \hspace{1pt} outperforms VERISUMM with 21 - 35\% gains in \textit{V-Ratio}. In terms of ROUGE-1-F1 scores, we outperform the strongest baseline SCC by a margin ranging between 2 - 5\% gains.}
    \label{tab:mtl-summarization-results}
    \resizebox{\linewidth}{!}{
    \begin{tabular}{|l|c|c|c|c|c|c|c|c|c|c|}
        \hline
        \multirow{2}{*}{\textbf{Model}} &
        \multicolumn{2}{c|}{\textbf{Charliehebdo}} &
        \multicolumn{2}{c|}{\textbf{Germanwings}} &
        \multicolumn{2}{c|}{\textbf{Ottawa}} &
        \multicolumn{2}{c|}{\textbf{Sydney}} &
        \multicolumn{2}{c|}{\textbf{Overall}} \\
        \cline{2-11}
        & \textbf{V-Ratio} & \textbf{R-1} 
        & \textbf{V-Ratio} & \textbf{R-1} 
        & \textbf{V-Ratio} & \textbf{R-1} 
        & \textbf{V-Ratio} & \textbf{R-1} 
        & \textbf{V-Ratio} & \textbf{R-1} \\
        \hline
        
        APSAL \cite{kedzie-acl-2015} & 0.200 & \textbf{0.493} & 0.056 & 0.437 & 0.000 & 0.435 & 0.000 & 0.483 & 0.064 & 0.462 \\
        \hline
        
        COWTS \cite{rudra2015extracting} & 0.385 & 0.479 & 0.111 & 0.504 & 0.118 & 0.427 & 0.087 & 0.479 & 0.175 & 0.472 \\
        \hline
        
        SCC \cite{rudra_subsum_2018} & 0.365 & 0.465 & 0.244 & 0.505 & 0.212 & 0.444 & 0.308 & 0.492 & 0.282 & 0.477 \\
        \hline
        
        VERISUMM \cite{sharma_verified_2019} & 0.722 & 0.381 & 0.624 & 0.519 & 0.645 & 0.396 & 0.619 & 0.378 & 0.653 & 0.419 \\
        \hline
        
        \quad w/ STLV & 0.781 & 0.352 & 0.703 & 0.555 & 0.720 & 0.362 & 0.730 & 0.417 & 0.734 & 0.422 \\
        \hline
        
        \multicolumn{11}{|l|}{\textbf{Our Methods}} \\
        \hline
        
        STLS & 0.667 & 0.430 & 0.789 & 0.510 & 0.629 & 0.431 & 0.632 & 0.474 & 0.679 & 0.461 \\
        \hline
        
        \quad + STLV & 0.749 & 0.454 & 0.825 & 0.558 & 0.732 & 0.416 & 0.732 & 0.487 & 0.760 & 0.479 \\
        \hline
        
        \multicolumn{11}{|l|}{\vsum} \\
        \hline
        
        \quad w/ $ \kappa = 1 $ & 0.750 & 0.444 & 0.927 & 0.565 & \textbf{0.860} & \textbf{0.477} & 0.627 & \textbf{0.518} & 0.791 & \textbf{0.501} \\
        \hline
        
        \quad w/ $ \kappa = 0.5 $ & 0.689 & 0.420 & 1.000 & \textbf{0.582} & 0.813 & 0.468 & 0.687 & 0.471 & 0.797 & 0.485 \\
        \hline
        
        \quad w/ $ \kappa = 0 $ & \textbf{0.947} & 0.472 & \textbf{1.000} & 0.575 & 0.834 & 0.464 & 0.750 & 0.452 & \textbf{0.883} & 0.491 \\
        \hline
        
    \end{tabular}
    }
\end{table*}
}

\newcommand{\tabmtlveriLoss}{
\begin{table*}
    \centering
    \caption{Comparing the Tweet Verification results of our model variants with those of the baselines. MTLV achieves 3.9\% \textit{Accuracy} gains and 1.5\% gains in \textit{Macro-F1} scores over VRoC. For \textit{Charliehebdo} and \textit{Ottawa}, the results of MTLV are statistically significant than STLV with 95\% confidence interval. Training the models with BERTweet further improves the results with MTLV achieving 4.5\% (\textit{Accuracy}) and 2.1\% (\textit{Macro-F1}) gains over VRoC. Top 2 scores in each column are highlighted.}
    \label{tab:mtl-verification-results}
    \resizebox{\textwidth}{!}{
    \begin{tabular}{|l|c|c|c|c|c|c|c|c|c|c|}
        \hline
        \multirow{2}{*}{\textbf{Model}} &
        \multicolumn{2}{c|}{\textbf{Charliehebdo}} &
        \multicolumn{2}{c|}{\textbf{Germanwings}} &
        \multicolumn{2}{c|}{\textbf{Ottawa}} &
        \multicolumn{2}{c|}{\textbf{Sydney}} & 
        \multicolumn{2}{c|}{\textbf{Overall}} \\
        \cline{2-11}
        & \textbf{Acc.} & \textbf{F1-Mac.} & \textbf{Acc.} & \textbf{F1-Mac.} & \textbf{Acc.} & \textbf{F1-Mac.} & \textbf{Acc.} & \textbf{F1-Mac.} & \textbf{Acc.} & \textbf{F1-Mac.} \\
        \hline
        
        
        
        CETM-TL \cite{sharma_verified_2019} & 0.788 & 0.737 & 0.644 & 0.621 & 0.748 & 0.748 & 0.727 & 0.710 & 0.727 & 0.704 \\
        \hline
        
        VRoC \cite{cheng_vroc_2020} & - & - & - & - & - & - & - & - & 0.752 & 0.752 \\ 
        \hline
        
        \multicolumn{11}{|l|}{\textbf{Our Methods}} \\
        \hline
        
        STLV & 0.821 & 0.770 & \textbf{0.742} & \textbf{0.741} & 0.766 & 0.766 & 0.759 & 0.748 & 0.772 & 0.756 \\
        \hline
        
        \quad + CETM & 0.822 & 0.770 & 0.723 & 0.722 & 0.765 & 0.764 & 0.759 & 0.751 & 0.767 & 0.752 \\
        \hline
        
        \quad w/o Tree-LSTM & 0.818 & 0.759 & 0.721 & 0.718 & 0.754 & 0.752 & 0.737 & 0.719 & 0.758 & 0.737 \\
        \hline
        
        w/ BERTweet & \textbf{0.841} & \textbf{0.783} & 0.732 & 0.726 & \textbf{0.780} & \textbf{0.780} & \textbf{0.779} & \textbf{0.773} & \textbf{0.783} & \textbf{0.766} \\
        \hline
        
        
        MTLV & 0.836 & 0.779 & \textbf{0.744} & \textbf{0.741} & 0.783 & 0.782 & 0.760 & 0.750 & 0.781 & 0.763 \\
        \hline
        
        w/ BERTweet & \textbf{0.849} & \textbf{0.788} & 0.729 & 0.721 & \textbf{0.785} & \textbf{0.785} & \textbf{0.781} & \textbf{0.779} & \textbf{0.786} & \textbf{0.768} \\
        \hline
        
        \textit{Mann-Whitney U test} & 0.045 & - & 0.094 & - & 0.041 & - & 0.089 & - & - & - \\
        \hline
        
    \end{tabular}
    }
\end{table*}
}

%% file: intro.tex
\vspace{-0.5em}
\section{Introduction}
\label{sec:intro}
Sudden adversities such as natural or man-made disasters escalate the speed of news sharing over social media in the form of posts containing critical situational updates, information about victims, and general public sentiments, among others \cite{Imran2020UsingAA}. Such platforms are however prone to widespread proliferation of misinformation, given their unmoderated nature, and the unavailability of sufficient updates from reliable sources to verify the credibility of messages at the time of posting \cite{zubiaga2018detection}. In such time-critical scenarios, it is nearly impossible for any concerned stakeholder, including humanitarian organizations, crisis-responders, and public in general, to manually fact-check such overwhelming volume of information \cite{ma2018rumor}. Motivated by these factors, and acknowledging the far-reaching detrimental consequences of falsehood \cite{Lazer2018TheSO}, we propose \textbf{\vsum}, an end-to-end solution that jointly \textbf{verifies} and \textbf{summarizes} large volumes of disaster-related tweets to generate information-rich and \textbf{trustworthy} summaries in near real-time.

Existing methods \cite{rudra2019summarizing, rudra_subsum_2018, rudra_crisis_2016, kedzie-acl-2015} to summarize \textit{disaster-specific} microblogs  do not make any explicit efforts to validate the authenticity of summary tweets, thereby making their solutions \textit{unreliable}. To the best of our knowledge, VERISUMM \cite{sharma_verified_2019} is the only proper baseline for the task of \textbf{verified/trustworthy summarization}. However, it takes a pipeline approach, and hence suffers from error propagation between the steps. We on the other hand \textbf{jointly} determine the \textit{credibility} and \textit{summary-worthiness} of a tweet by training \textit{\vsum} using a hierarchical multi-task learning setup. Such an approach intuitively mimics a pipeline while addressing its limitations. Prior \textit{disaster-specific} summarizers (including VERISUMM) are additionally \textit{unsupervised} in nature where the summaries are built by maximizing the presence of \textit{event-specific} keywords. It however affects their ability to generalize across unseen events, presumably with their own set of \textit{information classes} and associated keywords. To overcome this shortcoming, we design a \textbf{supervised} approach, as it can implicitly learn summary-worthy features along with the signals to identify a potentially \textit{rumourous} tweet when trained jointly with the \textit{verification} subtask.

Compared to the extensive literature on text summarization \cite{huang-etal-2020-achieved}, methods focusing on summarizing social media texts are limited \cite{Li_Zhang_2021}. In this work, we additionally make an attempt to address this gap by leveraging \textit{SummaRuNNer} \cite{nallapati_summaRunner}, one of the seminal works on extractive document summarization, and suitably modifying it to suit our purpose. More specifically, we need document-summary pairs to train the model, where documents consist of multiple sentences (tweets in our case). Here, instead of considering each tweet as a separate document or the entire tweet stream as a single document, we randomly select a subset of tweets to define a \textit{pseudo} document. However, by doing so, we restrict a tweet’s context to the document it is part of. We therefore allow the same tweet to be included in multiple such documents as long as each tweet is included in at least $m$ documents, where $m$ is a hyper-parameter. Finally, we propose a novel approach to approximate a tweet's global context from its multiple document-level local contexts.

Our end goal is to generate a summary with a high ratio of \textit{verified} or non-rumourous tweets to total tweets. The efficacy of our credibility \textit{verifier} is therefore of prime importance as it directly impacts the performance of the joint task. Twitter conversation threads possess a self-correcting nature as the true veracity of a claim gets revealed gradually through collective sense-making \cite{zubiaga2018detection}. Many prior works have tried to exploit this phenomenon \cite{kumar-carley-2019-tree, sharma_verified_2019, ma2018rumor, ma2017detect} by using tree-structured neutral networks such as Tree-LSTMs \cite{tai2015improved} to recursively capture the structural characteristics of such an information cascade. Ducci et. al. \cite{Ducci2020CascadeLSTMAT} recently proposed \textit{Cascade-LSTM}, a novel bi-directional variant of Tree-LSTM, to learn the \textit{complete} cascade. We extend it to additionally detect the \textit{stances} of each of the reply nodes towards the potentially rumourous source tweet \cite{zubiaga-etal-2016-stance}, to complete the design of our \textit{verifier} module. The social signals learnt, and the predictions made by the \textit{verifier} (trained at a lower layer) are used by the \textit{summarizer} (trained at a deeper layer) to determine the \textit{summary-worthiness} of a tweet. Tweet-level semantic features are learnt using BERTweet \cite{bertweet}, whose parameters are jointly updated by our multi-task objective. We use ILP to obtain the final summary by maximizing the \textit{relevance} scores of tweets thus obtained, while minimizing redundancy.

Our overall contributions can be summarized as follows:
\vspace{-0.5em}
\begin{itemize}[leftmargin=*]
    \item We present \vsum, the first end-to-end solution to obtain \textit{trustworthy summaries} from large volumes of disaster-related tweets. Using multi-task learning, \vsum \hspace{1pt} is trained to jointly determine the credibility and summary-worthiness of a tweet.
    \item Our approach is \textit{supervised}, which enhances the generalizability of our solution to unseen events. More importantly, we present a novel way to leverage an extractive document summarization technique for summarizing event (here, disaster)-specific tweets. 
    \item Our credibility \textit{verifier} improves upon the current state-of-the-art for the \textit{rumour detection} task by jointly learning the structural characteristics of information cascades, and stances of replies towards the source tweet. The features thus learnt help the \textit{verifier} and the \textit{summarizer} alike in respectively determining the veracity and summary-worthiness of source tweets.
    \item Extensive experiments on \textit{PHEME} \cite{zubiaga2015crowdsourcing} establish the better efficacy of \vsum \hspace{1pt} over strong state-of-the-art baselines. Comparative results along with the favourable outcomes of a qualitative human evaluation experiment are discussed in Section \ref{sec:results}.
\end{itemize}

We present the design of \vsum \hspace{1pt} in Section \ref{sec:model}. Our experimental setup, additional related work, and baselines are discussed in Section \ref{sec:experiments}. While our proposed solution is generic, we perform a thorough analysis of \vsum-generated summaries across several \textit{disaster-specific} classes relevant for PHEME events in Section \ref{subsec:class-specific}, and discover that the \textit{verified ratio} of summaries generated from \textit{Police Investigation}-related tweets substantially improves over time as authoritative sources gradually confirm or deny initially \textit{unverified} claims. We conclude our work in Section \ref{subsec:caseStudy} with a case study.

%% file: model.tex
\section{Methodology}
\label{sec:model}
\subsection{Problem Formulation}
\label{subsec:problem}
Let $T = \{T_1, T_2, ..., T_{\vert T \vert}\}$ represent a dataset of Twitter conversation threads posted during a crisis event, with $|T|$ representing the length of the dataset. Here, each thread $T_i = \{s_i, r_{i1}, r_{i2}, ..., r_{in}\}$ consists of a \textit{source} tweet or claim $s_i$ whose credibility needs to be verified. Further, we need to decide if $s_i$ should be included in the summary. The \textit{reply} tweets $r_{i*}$ are direct or indirect responses to $s_i$, thereby collectively forming a propagation tree like structure with $s_i$ as their root. Our multi-task objective is formulated as a supervised classification problem, which learns a classifier $f$ from labeled source tweets, $f: s_i \rightarrow \{Ver_i, Summ_i\}$, where $Ver_i \in \{\textit{verified}, \textit{unverified}\}$ and $Summ_i \in \{\textit{in-summary}, \textit{out-of-summary}\}$. We also capture the confidence scores $Ver_i^{prob}$ and $Summ_i^{prob}$ with which the final labels were predicted by the respective modules. These scores are used in the inference phase to generate the final summary.

\subsection{\vsum \hspace{1pt} Architecture}
\label{subsec:architecture}
\begin{figure}
    \centering
    \begin{tabular}{c}
        \includegraphics[width=0.7\columnwidth]{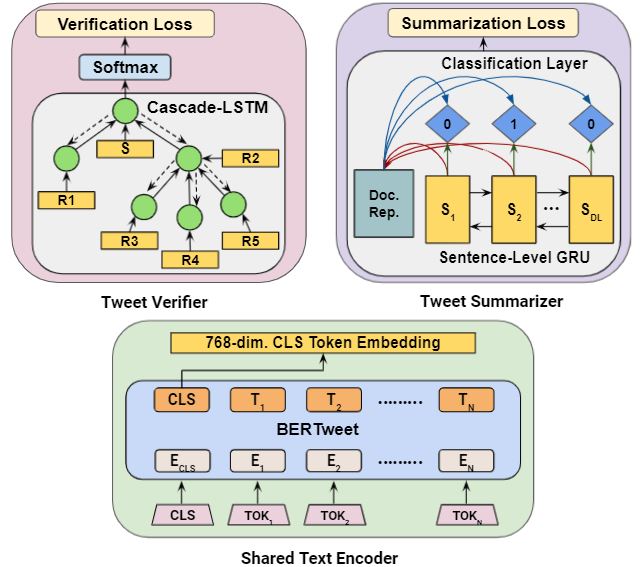}
        \\
        \includegraphics[width=0.75\columnwidth]{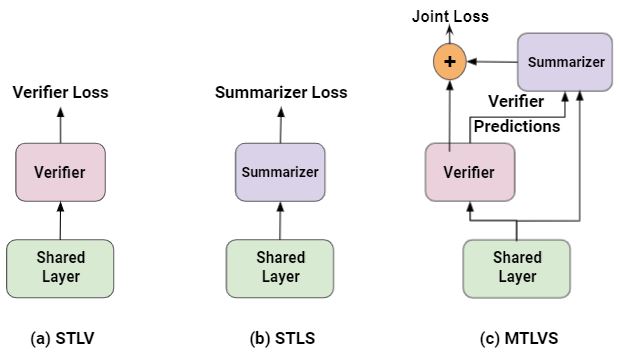}
        \\
    \end{tabular}
    \vspace{-1em}
    \caption{Dark yellow boxes represent the 768-dim. [CLS]-token embeddings of tweets. \textit{Tweet Verifier}: solid and dashed lines respectively represent the upward and downward Tree LSTM. Stance classification loss is obtained at each tree node. (a) STLV and (b) STLS: the \textit{verifier} and the \textit{summarizer} respectively trained as a single tasks, (c) \vsum: \textit{verifier} and \textit{summarizer} co-trained in a hierarchical multi-task setting.}
    \label{fig:model}
    \vspace{-1em}
\end{figure}

Fig. \ref{fig:model} illustrates the building blocks of our architecture, with Fig. \ref{fig:model}(c) showing the way they are connected in \vsum. We now detail below our model components and methodology for the joint task.


\subsubsection{\textbf{Shared Text Encoder}}
\label{subsubsec:sharedEncoder}
We use pre-trained BERTweet \cite{bertweet} as our text encoder. A given tweet is first tokenized by the \textit{BERTweet-Tokenizer}, prepended with the special $ [CLS] $ token, and passed through the encoder, which then learns rich contextualized representations for the tweet's constituent tokens. Given that both our modules are supervised classification tasks, we consider the 768-dimensional embedding of the $[CLS]$ token from the last layer of BERTweet as the final semantic representation of the tweet. BERTweet-parameters are respectively updated by the two modules when trained separately. For training \vsum, the text encoder is shared between the \textit{verifier} and the \textit{summarizer}, and the parameters are jointly fine-tuned by our multi-task objective, thereby facilitating interaction between the two modules.

\subsubsection{\textbf{Tweet Verification}}
\label{subsubsec:verifier}
Given a Twitter conversation thread $T_i$, the primary task of the credibility \textit{verifier} is to classify the source tweet $s_i$ as either \textit{verified}/non-rumourous or \textit{unverified}/rumourous. The secondary task is to classify each of the response tweets $r_{i*}$ as either \textit{supporting/ denying/ querying/ commenting on} the source tweet. We extend Cascade-LSTM \cite{Ducci2020CascadeLSTMAT} to jointly train both the tasks. Cascade-LSTM consists of two phases. In the first phase, an \textit{Upward Tree-LSTM} recursively processes the cascade tree via an upward traversal. As depicted in Fig. \ref{fig:model} (\textit{Tweet Verifier}), each node in the tree is passed through a Tree-LSTM unit taking two inputs: the $[CLS]$ token embedding of the corresponding tweet, and the bottom-up hidden state vectors ($h^{\uparrow}$) of its child nodes. The traversal terminates at the root ($s_i$), thereby updating its bottom-up hidden state vector ($h^{\uparrow}_{s_i}$) by accumulating the knowledge from the entire cascade.

In the second phase, a \textit{Downward Tree-LSTM} traverses the cascade from the root to the leaves, thereby learning the actual direction of the propagation dynamics. Now, the previously-held beliefs at each node are either reinforced or altered by each subsequent retweet/reply down the tree. During this phase, each reply node $r_{ij}$ is again passed through a Tree-LSTM unit taking two inputs: the corresponding tweet's $[CLS]$ token embedding, and the bottom-up hidden state vector of its parent to obtain its top-down hidden state vector, $h^{\downarrow}_{r_{ij}}$. The traversal terminates at the leaves. The LSTM gates, memory cell and hidden state at each node in both the phases are updated following the transition equations defined in \cite{Ducci2020CascadeLSTMAT}.

The concatenated bi-directional embedding $[h^{\uparrow}_{r_{ij}} : h^{\downarrow}_{r_{ij}}]$ at each response node $r_{ij}$ is passed through a feed-forward layer with softmax to obtain a normalized stance distribution over the four classes \textit{supporting}, \textit{denying}, \textit{querying}, and \textit{commenting}. For this task, we use \textit{cross-entropy} to define the loss function, $L_{stance}$. For the verification/rumour detection task, the concatenated feature vector $[h^{\uparrow}_{s_i} : h^{\downarrow}_{leaf} : [CLS]_{s_i}]$ is passed through a feed-forward layer with softmax at the root node of the tree to obtain the probability distributions for $s_i$ over the two classes, \textit{verified} and \textit{unverified}. Here, $h^{\downarrow}_{leaf}$ represents the average over the top-down hidden states of all leaves. \textit{Binary Cross-Entropy} is used to define the loss function, $L_{rumour}$. The \textit{verifier} is trained by optimizing the joint loss function, $L_{VER}$, with $\lambda_1$ being the trade-off parameter between the two losses.
\begin{equation}
    L_{VER} = L_{rumour} + \lambda_1 \cdot L_{stance}
\end{equation}


\subsubsection{\textbf{Tweet Summarization}}
\label{subsubsec:summarizer}
We model tweet summarization as a supervised classification task by leveraging \textit{SummaRuNNer} \cite{nallapati_summaRunner}. It consists of a two-layer bi-directional GRU-RNN. The first layer works at the \textit{word-level} to learn contextualized word representations which are then average-pooled to obtain sentence representations. We replace this layer by our BERTweet-based \textit{text encoder}, that is fine-tuned to obtain tweet semantic representations. The second layer of bi-directional GRU-RNN works at the \textit{sentence-level} (tweet-level in our case) to learn the contextualized representations of tweets in a document. Graphical representation of our modified network is depicted under \textit{Tweet Summarizer} module in Fig. \ref{fig:model}. 

We now define the notion of a \textit{document}. Let $S = \{ s_1, s_2, ..., s_{|T|} \}$ represent the set of all source tweets present in the set of tweet threads, $T$ with $|T|$ representing the dataset length. Our goal is to produce an extractive summary from $S$ while maximizing the \textit{verified ratio}, \textit{content-richness}, and \textit{novelty} of selected tweets, and minimizing redundancy. To efficiently train the \textit{summarizer}, we create multiple \textit{documents}, each consisting of a subset of available tweets. However by doing so, we restrict a tweet's context to the document it is part of. We explain the working of our model with this setup, before presenting a novel method to approximate the \textit{global} context of a tweet $s_i$ from its \textit{document-level} contexts. Let $D_k = \{s_1, s_2, ..., s_{DL} \}$ represent the $k^{th}$ document with length $DL$. All tweets in $D_k$ are sorted in ascending order of their posting time in order to learn any possible temporal clues. Following \cite{nallapati_summaRunner}, we obtain the document representation $d_k$ using the hidden states of the \textit{sentence-level} bi-directional GRU-RNN as follows:
\begin{equation}
    d_k = tanh (W_d \frac{1}{DL} \sum^{DL}_{i = 1} h_i^k + b)
\end{equation}
Here, $h_i^k$ represents the hidden state vector corresponding to $s_i \in D_k$. $W_d$ and $b$ respectively represent the weight and bias parameters. A \textit{classification layer} (Fig. \ref{fig:model}) then takes a binary decision regarding the summary-worthiness of $s_i$ by sequentially revisiting the tweets in a second pass. Probability ($p_i^{D_k}$) of $s_i$ to be included in the summary for document $D_k$ is obtained as:
\begin{equation}\label{eq:summ_decision}
    p_i^{D_k} = P (y_i^k = 1 \; | \; h_i^k, sum_i^k, d_k, Ver_i^{prob})
\end{equation}
The value of binary variable $y_i^k$ determines the inclusion/exclusion of tweet $s_i$ in the summary for $D_k$. Here, $sum_i^k$ represents the intermediate representation of the summary formed till $s_i$ is visited:
\begin{equation}
    sum_i^k = \sum_{j = 1}^{i - 1} h_j^k p_j^{D_k}
\end{equation}
Therefore, for a given tweet $s_i \in D_k$, its probability of being \textit{verified} ($Ver_i^{prob}$) as predicted by the underlying \textit{verifier}, its content ($h_i^k$), its importance given the context of the document ($d_k$), and possibility of adding redundant information considering the summary already formed ($sum_i^k$) are all taken into account while deciding upon its summary membership. We request our readers to refer to \cite{nallapati_summaRunner} for further details about the equations.

$p_i^{D_k}$ thus obtained is however limited to the context of $D_k$. Instead of considering $s_i$ only once, we therefore create additional documents randomly till each tweet $s_i$ gets included in at least $m = |T| / DL$ documents, where $m$ is a hyper-parameter. This helps us to approximate the summary-worthiness of $s_i$ with respect to a larger population of available tweets in $S$. For each tweet $s_i$, its overall summary-worthiness probability $Summ_i^{prob}$ is thus obtained as:
\begin{equation}\label{eq:summ_worthy_prob}
    Summ_i^{prob} = \frac{1}{m} \sum_{j=1}^m (p_i^{D_j} \; | \; s_i \in D_j)
\end{equation}
\textit{Binary Cross-Entropy} is used to define the \textit{summarizer} loss, $L_{SUMM}$:
\begin{equation}
    L_{SUMM} =  - \sum_{j=1}^N \sum_{i=1}^{DL} \left(y_i^{D_j} \cdot log(p_i^{D_j}) + (1 - y_i^{D_j}) \cdot log (1 - p_i^{D_j}) \right)
\end{equation}
Here, $N$ represents the total no. of documents created as detailed above. $y_i^{D_j} \in \{\textit{in-summary}, \textit{out-of-summary}\}$ represents the true membership of tweet $s_i$ in the summary for document ${D_j}$, and $p_i^{D_j}$ represents the corresponding label predicted by the \textit{summarizer}.

\subsubsection{\textbf{Joint Training}}
\label{subsubsec:training}
We minimize the following joint loss to co-train the two modules in a hierarchical multi-task learning setup:
\begin{equation}\label{eq:mtl_obj}
    L_{\vsum} = L_{VER} + \lambda_2 \cdot L_{SUMM} 
\end{equation}
Here, $\lambda_2$ represents a trade-off between the two losses. The \textit{verifier} is trained in batches of tweets, whereas the \textit{summarizer} is trained in batches of documents. Each training epoch is therefore divided into two steps. First, we train the \textit{verifier} at a lower layer to obtain the probability score ($Ver_i^{prob}$) of source tweet $s_i$ being non-rumourous. In the second step, these scores are used (Eq. \ref{eq:summ_decision}), while training the \textit{summarizer} at a deeper layer, to decide the summary-worthiness of tweets. As the joint loss backpropagates through the \textit{shared encoder}, BERTweet parameters are fine-tuned end-to-end to produce latent representations of tweets which benefit both the tasks.

\subsubsection{\textbf{Scoring the Tweets}}
\label{subsubsec:relevance_scoring}
For scoring, we consider the \textit{verifier}-generated probability score $Ver_i^{prob}$, and the \textit{summarizer}-generated probability score $Summ_i^{prob}$ (given by Eq. \ref{eq:summ_worthy_prob}), for all those source tweets $s_i$'s where both the scores are greater than 0.5. We term these source tweets as \textit{relevant}. Here we postulate that including tweets with high $Summ^{prob}$ scores in our generated summaries would ideally help in achieving high ROUGE1-F1 scores with respect to the gold-standard summaries, thereby ensuring high \textit{content-richness}; whereas including tweets with high $Ver^{prob}$ scores would help in improving the \textit{verified-ratio} of summary tweets. In order to allow for this trade-off, we introduce a balancing parameter $ \kappa $ to define the \textbf{relevance-score} of the source tweet $s_i$ as follows:
\begin{equation}\label{eq:relScore}
    Score(s_i) = \kappa \cdot Summ_i^{prob} + (1 - \kappa) \cdot Ver_i^{prob}
\end{equation}

\subsubsection{\textbf{Inference - Summary Generation}}
\label{subsubsec:inference}
For generating the final summary, our goal is to maximize the collective \textit{relevance-scores} of the selected subset of tweets while minimizing redundancy. Let the desired length of summary be $ L $ words. We formalize these objectives as a constrained maximization problem using ILP. In the following equations, let $l_i $ be the length (in words) of the source tweet $s_i$. Let $x_i$ be a binary indicator variable whose value indicates whether to include $s_i$ in the final summary. $sim_{ij}$ represents the cosine similarity between the embeddings (obtained using \textit{Sentence-Transformers} \cite{reimers-2019-sentence-bert}) of tweets $ s_i $ and $s_j$. We use the \textit{Gurobi optimizer} \cite{gurobi} to maximize the objective as defined in Eq. \ref{eq:ILP}:
\begin{equation}\label{eq:ILP}
{\operatorname{argmax}} \; \sum_{i} Score(s_i) \ x_i \ - \sum_{ij} sim_{ij} \ y_{ij}
\end{equation}
\begin{equation}\label{Length_Constraint}
s.t. \; \; \sum_{i} l_i x_i \; \leq \; L
\end{equation}
\begin{equation}\label{Redundancy_Constraint1}
\; \; y_{ij} \; \leq \; \frac{1}{2} (x_i + x_j ) \; \; \forall \ i,j
\end{equation}
\begin{equation}\label{Redundancy_Constraint2}
\; \; y_{ij} \; \geq \; x_i + x_j -1 \; \; \forall \ i,j
\end{equation}
Eq. \ref{Length_Constraint} restricts the length of the summary to $ L $ words. Eqs. \ref{Redundancy_Constraint1} and \ref{Redundancy_Constraint2} ensure that $y_{ij}$, another binary variable, is $1$ if and only if both $x_i$ and $x_j$ are $1$. Therefore, apart from maximizing the collective \textit{relevance} of the selected subset of tweets, Eq. \ref{eq:ILP} tries to minimize their collective mutual similarity scores, thereby minimizing \textit{redundancy} while enhancing the \textit{content-richness} and the \textit{novelty} of the information being covered. We therefore note that the final summary-membership of each \textit{relevant} source tweet $s_i$ is obtained considering the global context of all relevant candidate tweets in $S$.

%% file: experiments.tex
\section{Experiments}
\label{sec:experiments}

\subsection{Dataset and Annotation}
\label{subsec:dataset}
We use the \textbf{PHEME} dataset \cite{zubiaga2015crowdsourcing} consisting of $ 4,659 $ Twitter conversation threads, each with a source tweet and its replies, posted during four breaking-news events related to man-made disasters. For each thread, the source tweet is labeled as either \textit{rumour} or \textit{non-rumour} by a team of journalists. We consider the non-rumourous tweets as \textit{verified} and rumourous tweets as \textit{unverified}. Stance labels for a subset of \textit{rumourous} threads were obtained from the \textit{RumourEval 2019} dataset \cite{gorrell-etal-2019-semeval}. Table \ref{tab:pheme5-satistics} contains the dataset statistics. 

For each of the four events, we obtain experts-curated extractive summaries (one per event) from Sharma et. al. \cite{sharma_verified_2019}, all of which are around 250 words long. It is to be noted here that both the gold-standard as well as our model-generated summaries consist only of source tweets (and not replies) since the rumour/non-rumour annotations are only defined for the source tweets. Furthermore, all gold-standard summary tweets are  non-rumourous/\textit{verified} as per PHEME annotations. For each event, we thus have only around 20 tweets labeled as \textit{in-summary} thereby making the dataset too skewed (towards the \textit{out-of-summary} class) to efficiently train a supervised classifier for the \textit{summarization} task. We therefore consider changing the labels of a few \textit{out-of-summary} tweets (\textit{verified} as per PHEME) to \textit{in-summary}, based upon the cosine similarities of their embeddings with those of the original \textit{in-summary} tweets. Tweet embeddings are obtained using \textit{Sentence-Transformers} \cite{reimers-2019-sentence-bert}, a state-of-the-art approach for producing semantically meaningful sentence embeddings. The similarity threshold was empirically set in the range of [0.7, 0.8] so that we could obtain a ratio of nearly 1:9 between \textit{in-summary} and \textit{out-of-summary} tweets for each of the events. While these updated labels are used \textbf{only for training}, ROUGE1-F1 scores of model-generated summaries are calculated with respect to the \textbf{original gold-standard} \textit{in-summary} tweets.

\subsection{Experimental Setup}
\label{subsec:setup}
\tabdataset
We use the \textit{bertweet-base} pre-trained model from the \textit{HuggingFace} library to initialize the parameters of our \textit{shared encoder}. In the \textit{verifier}, hidden state dimensions of the Tree-LSTM cells are set to 128. As illustrated in \cite{sharma_verified_2019}, 80-90\% of the replies are at levels $\leq 5$. Hence, during training, the maximum height of propagation trees is set to 5. We note here however that our trained model can be safely applied to new events for determining the credibility (and summary-worthiness) of source tweets in near real-time without having to wait for their replies. Our codes are suitably adjusted to handle such cases. For training our \textit{summarizer}, we use the default parameters as defined in \cite{nallapati_summaRunner}. Other model parameters are randomly initialized.


All our models, including the baselines, are trained using the \textbf{Leave-one-out} ($ L $) principle, where in each run of the experiment, the data from one event is set-aside as test-set while the labeled tweets from all other events are used for training. This helps us to evaluate and compare the generalizability of our proposed solution. We set aside 10\% of the training data as validation set for hyper-parameter tuning. While doing so, we maintain similar class distributions in the \textit{train} and \textit{val} sets for the corresponding tasks. For each train-test combination, we perform extensive experiments with different sets of hyper-parameters, select the model corresponding to the best validation performance, and evaluate it on the test set. Our final models are trained with learning rates ranging from $ 1e-6 $ to $ 1e -4 $, optimizer weight decays ranging from $ 0.1 $ to $ 0.001 $, and no. of epochs ranging from $ 5 $ to $ 20 $. All our models are trained end-to-end with \textit{Adam} optimizer \if 0\cite{Kingma2015AdamAM}\fi on Tesla P100-PCIE (16GB) GPU. Finally, we ensure the reproducibility of our results and make our codes and preprocessed datasets publicly available \footnote{\url{https://github.com/rajdeep345/MTLTS}}.

\subsection{Evaluation Metrics}
\label{subsec:metrics}
From Table \ref{tab:pheme5-satistics}, we observe a considerable imbalance in \textit{verified} and \textit{unverified} classes in our dataset. Hence, for evaluating the rumour detection/\textit{verification} performance, we use the macro-averaged F1 (or \textbf{Macro-F1}) scores. We however also report the \textbf{Accuracy} scores following prior works. For the evaluation of model-generated summaries, we consider the following metrics: (i) \textbf{V-Ratio} (\textit{verified ratio}): representing the proportion of \textit{verified} tweets in the summaries, and (ii) \textbf{R-1}: representing the ROUGE-1 F1 scores obtained against the gold-standard summaries (all 250 words long). 

\subsection{Related Work and Baselines}
\label{subsec:baselines}
\subsection*{Tweet Verification}
\label{subsubsec:verification_baselines}
Recent research on detecting rumours from social media conversations predominantly relies on tree-based neural classifiers \cite{ma2018rumor, ma2017detect} to learn the temporal and structural properties of information cascades. \cite{kumar-carley-2019-tree, ma2018detect} additionally exploit the stances of community responses to learn rumour-specific signals. Different from these, multi-task learning has been used in \cite{cheng_vroc_2020, kochkina2018all} to train several rumour-related subtasks. Recently, \textit{Cascade-LSTM} (Ducci et. al. \cite{Ducci2020CascadeLSTMAT}) and Min et. al. \cite{Khoo2020InterpretableRD} have addressed the limitations of tree-based approaches. Our \textit{verifier} (Section \ref{subsubsec:verifier}) combines the advantages of \textit{Cascade-LSTM} and \textit{stance-classification} to eventually improve the \textit{rumour detection} performance. We compare the performance of \textbf{STLV} (Fig. \ref{fig:model}a) with the following representative state-of-the-art baselines: 
\begin{itemize}[leftmargin=*]
    \item \textbf{CETM-TL} (Sharma et al. \cite{sharma_verified_2019}): a bottom-up Tree LSTM-based classifier that leverages tweet features derived from an unsupervised LDA-based Content Expression Topic Model (CETM). 
    \item \textbf{TD-RvNN} (Ma et al. \cite{ma2018rumor}): a top-down \textit{Recursive Neural Networks}-based rumour classifier.
    \item \textbf{TL-Conv} (Kumar et al. \cite{kumar-carley-2019-tree}: Tree LSTM with convolution units to jointly address rumour detection and stance classification.
    \item \textbf{VRoC} (Cheng et al. \cite{cheng_vroc_2020}): a variational autoencoder-based rumour classifier that jointly trains all four components of a rumour classification system as defined in \cite{zubiaga2018detection}. For VRoC, we consider the results of its \textit{rumour detection} component for our comparisons.
    \item \textbf{Cascade-LSTM} (Ducci et al. \cite{Ducci2020CascadeLSTMAT}: proposes a novel bi-directional variant of Tree-LSTM to accumulate the knowledge from the entire information cascade at each tree node.
\end{itemize}

\vspace{-0.2cm}
\subsection*{Trustworthy Summarization}
\label{subsubsec:summary_baselines}
\tabmtlsummRichness

Studies on summarizing social media data are limited in comparison to the vast literature on document summarization \cite{Li_Zhang_2021}. Addressing this gap was one of our prime motivations. Accordingly, we leverage \textit{SummaRuNNer} \cite{nallapati_summaRunner}, and present a novel way to make it suitable for summarizing disaster-related tweets. Exploring the possibility of utilizing more recent extractive document summarization techniques for Twitter summarization is envisaged as a future work. Given the focus of our study, we compare the performance of \vsum \hspace{1pt} with the following \textit{disaster-specific} summarizers:
\begin{itemize}[leftmargin=*]
    \item \textbf{APSAL} (Kedzie et al. \cite{kedzie-acl-2015}) is a clustering-based multi-document summarization system that focuses on human-generated information nuggets to assign salience scores to disaster-related articles.
    \item \textbf{COWTS} (Rudra et al. \cite{rudra2015extracting}) is an extractive summarization approach specifically designed to generate situational summaries from disaster-related tweets.
    \item \textbf{SCC} (Rudra et al. \cite{rudra_subsum_2018}) is an extractive summarization approach that first identifies sub-events from disaster-related tweets, followed by obtaining a summary for each identified sub-event. Taking cues from Sharma et al \cite{sharma_verified_2019}, we identify four sub-events relevant to PHEME events (refer Section \ref{subsec:class-specific}), obtain a summary for each sub-event, and report the average scores for comparison.
    \item \textbf{VERISUMM} (Sharma et al. \cite{sharma_verified_2019}) is the existing state-of-the-art for this task. It takes a pipeline approach by first verifying the tweets using CETM-TL, followed by solving a constrained-maximization problem (using ILP) for jointly optimizing the credibility and content richness of the generated summaries. 
\end{itemize}

\noindent We further perform a few \textbf{ablations} of our proposed architecture: 
\begin{itemize}[leftmargin=*]
    \item \textbf{STLS}, as depicted in Fig. \ref{fig:model}b, represents the \textit{summarizer} when trained as a single task (without tweet credibility verification).
    \item \textbf{STLS + STLV}: We train STLS and STLV separately for each train-test combination. During inference, we use the tweet-level classifier probabilities, respectively generated by the two modules, for generating the final summaries.
\end{itemize}

%% file: results.tex
\section{Main Results}
\label{sec:results}
\begin{table}
    \centering
    \caption{Qualitative comparison of Rumour Detection/Tweet Verification results, averaged across all four train/test splits over the PHEME dataset.}
    \label{tab:stl-verification-results}
    \resizebox{\columnwidth}{!}{
    \begin{tabular}{|l|c|c|c|c|}
        \hline
        \textbf{Model} & \textbf{Accuracy} & \textbf{Precision} & \textbf{Recall} & \textbf{F1-Macro} \\
        \hline
        CETM-TL \cite{sharma_verified_2019} & 0.727 & 0.715 & 0.693 & 0.704 \\
        \hline
        TD-RvNN \cite{ma2018rumor} & 0.737 & 0.748 & 0.738 & 0.743 \\
        \hline
        TL-Conv \cite{kumar-carley-2019-tree} & 0.740 & 0.743 & 0.747 & 0.745 \\
        \hline
        VRoC \cite{cheng_vroc_2020} & 0.752 & 0.755 & 0.752 & 0.752 \\
        \hline
        Cascade-LSTM \cite{Ducci2020CascadeLSTMAT} & 0.768 & 0.762 & 0.756 & 0.759 \\
        \hline
        \hline
        STLV & 0.783 & 0.767 & 0.765 & 0.766 \\
        \hline
        MTLV & \textbf{0.786} & \textbf{0.770} & \textbf{0.766} & \textbf{0.768} \\
        \hline
        \hline
        \textit{Mann-Whitney U test} & 0.039 & - & - & - \\
        \hline
    \end{tabular}
    }
    \vspace{-1em}
\end{table}

\subsection{Evaluating \vsum}
\label{subsec:results_mtlvs}
We compare the results of \vsum \hspace{1pt} with our \textit{summarizer} ablations and baselines for the joint task of \textit{trustworthy summarization} in Table \ref{tab:mtl-summarization-results}. Here, we remind our readers that $\kappa$ represents a balancing parameter that allows for a trade-off between the \textit{verified ratio} (or \textit{V-Ratio}) and \textit{content-richness} of generated summaries (refer Eq. \ref{eq:relScore}). Therefore, for $\kappa = 1$, we expect better content-richness in terms of high \textit{ROUGE-1 F1} scores, while for $ \kappa = 0 $, we expect high \textit{V-Ratio} scores. For \vsum, overall we observe this trend which confirms our hypothesis. Here, we reiterate that improving the \textit{V-Ratio} of generated summaries remains our primary objective.

We note that for APSAL, COWTS, and SCC, the \textit{V-Ratio} scores are very poor, whereas their ROUGE-1 F1 scores are fairly high. This is because these methods were designed to optimize the content-richness of generated summaries. In terms of \textit{V-Ratio}, VERISUMM outperforms the above three methods since it was optimized for the joint task. Interestingly, we find that STLS, despite receiving no external cues from the \textit{verifier} as against VERISUMM, outperforms the latter with an overall 4\% gain in \textit{V-Ratio} scores. This observation may be attributed to the better learning capabilities of a supervised task (STLS) over an unsupervised approach taken in VERISUMM. We further observe that all the variants of \vsum \hspace{1pt} attain very high \textit{V-Ratio} scores. This is owing to the fact that during co-training, the \textit{summarizer} utilizes the underlying \textit{verifier} predictions while determining the summary-worthiness of candidate tweets. To conclude, with overall \textbf{21 - 35\%} gains in \textit{V-Ratio} scores over VERISUMM, and with \textbf{2 - 5\% (16 - 20 \%)} gains in \textit{ROUGE-1 F1} scores over SCC (VERISUMM), \vsum \hspace{1pt} achieves new state-of-the-art on the PHEME dataset for the task of \textit{trustworthy summarization}. These observations establish the better efficacy of \vsum \hspace{1pt} in generating highly trustworthy and content-rich summaries from crisis-related tweets.

\subsection{Evaluating Tweet Verification}
\label{subsec:veri_comparison}
All methods (including \textit{STLV}) are trained using the $L$ principle (Section \ref{subsec:setup}) for each train-test combination of PHEME events. Scores averaged over four runs of the experiment (each with one of the four events as the test set) are reported in Table \ref{tab:stl-verification-results}. Results for \textit{VRoC} are however taken from Cheng et al.  \cite{cheng_vroc_2020}. \textbf{MTLV} represents the verification scores obtained during the training of \vsum.

From Table \ref{tab:stl-verification-results}, we note that TD-RvNN comfortably outperforms CETM-TL, thereby demonstrating the better learning capabilities of a top-down tree-based classifier. TL-Conv, despite being a bottom-up approach, marginally outperforms TD-RvNN, which demonstrates the advantage of jointly learning stance detection with rumour classification. VRoC performs better than the previous three methods since its \textit{rumour detection} component is benefited from the simultaneous training of three other related tasks. The results of Cascade-LSTM well-establish the importance of learning bi-directional embeddings at each tree node. STLV combines the best of both worlds by extending the architecture of Cascade-LSTM to jointly train stance classification at each of the response nodes in the message propagation tree. We achieve new state-of-the-art performance on the PHEME dataset for the task of rumour detection, with our results being statistically significant than Cascade-LSTM with 95\% confidence interval. MTLV results are comparable with those of STLV. Here, we would like to highlight that \textit{tweet verification} plays the role of an auxiliary task in our case for eventually improving the performance of \textit{trustworthy summarization}.

%% file: analysis.tex
\section{Analysis and Discussions}
\label{sec:analysis}
\begin{table}
	\centering
	\caption{Human-Eval. Results with 70 respondents. Values indicate (\%) times a method is preferred for a given question.}
	\label{tab:survey_results}
	\resizebox{\columnwidth}{!}{
	\begin{tabular}{|l|l|l|l|l|l|l|l|l|l|l|l|l|}
		\hline
		\multicolumn{1}{|c|}{\multirow{2}{*}{\textbf{Methods}}} & \multicolumn{3}{c|}{\textbf{Charliehebdo}} & \multicolumn{3}{c|}{\textbf{Germanwings}} & \multicolumn{3}{c|}{\textbf{Ottawa}} & \multicolumn{3}{c|}{\textbf{Sydney}} \\ 
		\cline{2-13} 
		\multicolumn{1}{|c|}{} & \multicolumn{1}{c|}{\textbf{Q1}} & \multicolumn{1}{c|}{\textbf{Q2}} & \multicolumn{1}{c|}{\textbf{Q3}} & \multicolumn{1}{c|}{\textbf{Q1}} & \multicolumn{1}{c|}{\textbf{Q2}} & \multicolumn{1}{c|}{\textbf{Q3}} & \multicolumn{1}{c|}{\textbf{Q1}} & \multicolumn{1}{c|}{\textbf{Q2}} & \multicolumn{1}{c|}{\textbf{Q3}} & \multicolumn{1}{c|}{\textbf{Q1}} & \multicolumn{1}{c|}{\textbf{Q2}} & \multicolumn{1}{c|}{\textbf{Q3}} \\ 
		\hline
		COWTS & 0.07 & 0.14 & 0.37 & 0.11 & 0.14 & 0.10 & 0.24 & 0.13 & 0.23 & 0.13 & 0.08 & \textbf{0.41} \\ 
		\hline
		
		SCC & 0.16 & 0.07 & 0.10 & 0.23 & 0.10 & 0.27 & 0.10 & 0.10 & 0.29 & 0.27 & 0.23 & 0.09 \\ 
		\hline
		
		VERISUMM & 0.23 & 0.16 & 0.14 & 0.10 & 0.20 & 0.16 & 0.16 & 0.21 & 0.11 & 0.13 & 0.13 & 0.13  \\ 
		\hline
		
		\textbf{\vsum} & \textbf{0.54} & \textbf{0.63} & \textbf{0.39} & \textbf{0.56} & \textbf{0.56} & \textbf{0.47} & \textbf{0.50} & \textbf{0.56} & \textbf{0.37} & \textbf{0.47} & \textbf{0.56} & 0.37 \\ 
		\hline
	\end{tabular}
	}
	\vspace{-0.5em}
\end{table}

\subsection{\textbf{Why Multi-task Learning?}}
\label{subsec:motivation}
In order to understand the scope of taking a multi-task approach, we first train the \textit{verifier} and \textit{summarizer} modules separately and report our results for STLS and STLV in Tables \ref{tab:mtl-summarization-results} and \ref{tab:stl-verification-results} respectively. STLS results highlight a considerable scope of improvement in the \textit{V-Ratio} scores. For each train-test combination, we then explicitly use the \textit{verifier} predictions to calculate the \textit{relevance-scores} of tweets (following Eq. \ref{eq:relScore}, with $\kappa = 0.5$), before reconstructing the summaries using ILP. Considerable (overall 11.9\%) gains in \textit{V-Ratio} scores (row STLS + STLV in Table \ref{tab:mtl-summarization-results}) prompted us to explore the advantage of training them jointly using a hierarchical multi-task setup, with \textit{trustworthy summarization} being the main task, and \textit{tweet verification} playing the role of an auxiliary task.


\subsection{\textbf{Human Evaluation Results}}
We perform a qualitative human evaluation of \vsum-generated summaries ($\kappa = 0$) against our baselines (all around 250 words long). Specifically, we created a survey that was divided into four sections corresponding to the four events under consideration. In each section, we randomly placed the summaries generated by four different systems, i.e. COWTS, SCC, VERISUMM, and \vsum \hspace{1pt} without revealing their identities. Details can be found in our code \href{https://github.com/rajdeep345/MTLTS}{\textit{repository}}. Participants were then asked to evaluate the summaries in each section based on three questions: \textbf{Q1 (diversity)} \textit{Which of the summaries contains the least amount of redundant information}? \textbf{Q2 (reliability)} \textit{Which of the summaries contains the most no. of reliable  tweets}?, and \textbf{Q3 (situational awareness)} \textit{Which of the summaries contains the most no. of situational tweets}? A total of 70 grad students completed the entire survey. For each question corresponding to each event, we report in Table \ref{tab:survey_results} the fraction of times a particular method was chosen over the others. With regards to capturing verified and less redundant information, \vsum\ comfortably outperforms VERISUMM. Although COWTS summaries for \textit{Sydney Siege} are marginally better in terms of capturing high proportion of situational content, they get substantially outclassed by \vsum\ summaries for \textit{Germanwings Crash}, and \textit{Ottawa Shootings}. Overall, the respondents found the \vsum-generated summaries to be more trustworthy, actionable and with less redundant information.

\subsection{\vsum \hspace{1pt} Summaries and Information Classes Specific to Man-made Disasters}\label{subsec:class-specific}
\begin{table}
    \centering
    \caption{Categories of crisis-related tweets in the PHEME dataset along with their corresponding Keywords.}
    \label{tab:classWiseKeywords}
    \resizebox{\columnwidth}{!}{
    \begin{tabular}{|p{1.8cm}|p{7.5cm}|p{7.5cm}|}
        \hline
        \textbf{Class} & \multicolumn{1}{c|}{\textbf{Keywords obtained using CETM}} & \multicolumn{1}{c|}{\textbf{Keywords obtained using CatE}} \\
        \hline
        
        \textbf{Victims and Casualties} & people, victims, families, friends, condolences, attack, lives, artist, killed, murder, country & family, loved, murdered, artists, tignous, cabu, wolinski, charb, stephane, charbonnier, colleagues, relatives, crew\\
        \hline
        
        \textbf{Investigation} & police, shooting, shot, killed, gunman, suspect, video, confirm, police, armed, cops, photo, gas, forces, justice & uniformed, ammunition, alleged, investigating, michael, zehaf, sergeant, security, soldier, kevin, vickers, official \\
        \hline
        
        \textbf{Other Event-Specific} & place, authority, supermarket, blocked, pilot, cockpit, locked, cafe, siege, hostages, scene, site, bystanders & witness, hostages\_flee, selfie, lindt, cafe, grocery, opera\_house, porte, vincennes, shop, signal, rescue  \\
        \hline
        
        \textbf{Opinions} & people, thoughts, muslim, hostage, condolences, family, thinking, sympathy, justice, sydney, ottawa, situation & horrible, scary, blamed, saddened, sorry, heart, terrible, really, wish, hope, thinking, sending, staysafeottawa \\
        \hline
    \end{tabular}
    }
    \vspace{-0.5em}
\end{table}

Different stakeholders might be interested in different aspects of the ongoing crisis, such as \textit{information about victims, updates on police investigations, regions affected by the event}, and general \textit{public sentiments}, among others. Hence, it becomes crucial to analyze the extent to which \vsum\ summaries capture trustworthy content across different classes of information specific to man-made disasters. We therefore need to first identify such classes relevant for the PHEME events. Based upon our observations and taking cues from \cite{sharma_verified_2019}, we consider the following four classes: (a) \textbf{Victims and Casualties}, (b) \textbf{Police Investigation}, (c) \textbf{Other Event-specific Updates}, and (d) \textbf{Personal Opinions}. In the absence of gold-standard labels, we use \textit{WeSTClass} \cite{westClass}, a keyword-driven text classifier to predict the information class for each tweet. Performance of such a classifier however depends upon the discriminative nature of the embedding space it works with and the quality of weak-supervision provided to it by means of seeds/keywords for the expected categories/classes. We explore two methods here.

We first examine the top words obtained using the \textit{Content Expression Topic Model} (CETM) \cite{sharma_verified_2019} for each of the information classes identified above. From Table \ref{tab:classWiseKeywords}, we notice the existence of overlapping keywords across classes, for e.g., words such as ``condolences" and ``family" appear under both \textit{Victims and Casualties}, and \textit{Public Opinions}. This known limitation of such unsupervised LDA-based topic models makes the downstream text classification task difficult. We therefore leverage \textit{CatE} \cite{catE}, a recently proposed category-name guided text embedding method, that takes a set of category/class names as user guidance (with optional seed words for each class) to learn category distinctive word embeddings which help in retrieving a set of representative and discriminative terms under each given category. We train \textit{CatE} with the information class names identified above along with their corresponding keywords obtained using CETM, and report the top keywords retrieved by \textit{CatE} for each class in Table \ref{tab:classWiseKeywords}. The keywords are noticeably {\bf distinctive and semantically consistent} with their corresponding class names (for e.g., names of victims and suspects being mapped to respective classes). We therefore continue with \textit{CatE} for further analysis.

\begin{table}
    \centering
    \caption{Distribution of tweets across Information Classes specific to PHEME events.}
    \label{tab:classWiseTweetDist}
    \resizebox{\columnwidth}{!}{
    \begin{tabular}{|l|c|c|c|c|c|c|c|c|}
        \hline
        \multirow{2}{*}{\textbf{Event}} & \multicolumn{2}{c|}{\textbf{Victims}} & \multicolumn{2}{c|}{\textbf{Investigation}} & \multicolumn{2}{c|}{\textbf{Other Event-Specific}} & \multicolumn{2}{c|}{\textbf{Opinions}} \\
        \cline{2-9}
        & \textbf{Ver.} & \textbf{UnVer.} & \textbf{Ver.} & \textbf{UnVer.} & \textbf{Ver.} & \textbf{UnVer.} & \textbf{Ver.} & \textbf{UnVer.} \\
        \hline
        
        Charliehebdo & 768 & 82 & 274 & 349 & 22 & 10 & 557 & 17 \\
        \hline
        
        Germanwings & 59 & 3 & 1 & 1 & 132 & 218 & 39 & 16 \\
        \hline
        
        Ottawa & 86 & 26 & 102 & 377 & 21 & 12 & 211 & 55\\
        \hline
        
        Sydney & 64 & 4 & 40 & 69 & 178 & 370 & 417 & 79 \\
        \hline
        \hline
        
        \textbf{Total} & \textbf{977} & \textbf{115} & \textbf{417} & \textbf{796} & \textbf{353} & \textbf{610} & \textbf{1224} & \textbf{167} \\
        \hline
    \end{tabular}
    }
\end{table}

The discriminative word embedding space learnt using \textit{CatE} is then used to train the \textit{WestClass} text classifier with the same information class names provided to it as weak supervision. Distribution of \textit{verified} and \textit{unverified} tweets (as per PHEME annotations) across all four information classes (using \textit{WestClass} predictions) are reported in Table \ref{tab:classWiseTweetDist}. We find that around 30\% of the tweets (1,391 out of total 4,659) contain public opinions or \textit{non-situational} content, out of which around 87\% are \textit{verified}. This observation led us to investigate if the improved \textit{V-Ratio} scores of \vsum\ summaries were owing to the inclusion of majorly non-situational tweets. We therefore generate the summaries for all four events with three separate values of $\kappa (0, 0.5, 1)$, and report the percentage distribution of summary tweets across the four information classes in Table \ref{tab:classWiseSummTweetDist}. Averaged over all summaries, only around 15\% of the summary tweets were found to be non-situational, thereby leading us to the conclusion that in addition to capturing trustworthy information, \vsum-generated summaries contain substantial amount of {\bf actionable} or {\bf situational} content. 


Lastly, we examine the \textit{verified ratio} of \vsum\ summaries, generated separately for each class of information. Since rumorous claims posted during emergencies are eventually either confirmed or denied by reliable sources, we additionally investigate the \textit{eventual verified ratio} (EVR) of our class-wise summaries using the \textit{veracity} labels of tweets obtained from a subsequent version \cite{pheme9} of PHEME. We discover from Table \ref{tab:classWiseVer} that many tweets \textbf{originally marked as unverified become eventually verified} across all classes. The summaries generated from tweets related to \textit{Police Investigations} are especially less reliable in the beginning, despite containing potentially true information. As official details are gradually released by authoritative agencies, originally \textit{unverified} tweets change their \textit{veracity} status leading to a substantial improvement in \textit{verified ratio} (\textit{EVR} scores) for the same set of summaries for the class.

\begin{table}
    \centering
    \caption{Percentage distribution of \vsum\ summary tweets across Information Classes specific to PHEME events. Values in brackets correspond to the \%-age of tweets for a given class in the summaries for (Charlie Hebdo, Germanwings crash, Ottawa shooting, Sydney Siege).}
    \label{tab:classWiseSummTweetDist}
    \resizebox{\columnwidth}{!}{
    \begin{tabular}{|p{1.5cm}|c|c|c|c|}
        \hline
        \textbf{Balancing Parameter} & \textbf{Victims} & \textbf{Investigation} & \textbf{Other Event-Specific} & \textbf{Opinions} \\
        \hline
        
        $\kappa = 0 $ & 38.5\% (67, 69, 6, 12) & 23.5\% (22, 0, 72, 0) & 17.75\% (0, 25, 11, 35) & 20.25\% (11, 6, 11, 53) \\
        \hline
        
        $\kappa = 0.5 $ & 18\% (12, 50, 5, 5) & 37.25\% (70, 0, 79, 0) & 32\% (12, 44, 5, 67) & 12.75\% (6, 6, 11, 28) \\
        \hline
        
        $\kappa = 1 $ & 17.5\% (0, 59, 6, 5) & 39.25\% (82, 0, 70, 5) & 32.25\% (12, 41, 6, 70) & 11\% (6, 0, 18, 20) \\
        \hline
        \hline
        
        \textbf{Overall} & \textbf{24.67\%} (26.3, 59.3, 5.7, 7.3) & \textbf{33.33\%} (58, 0, 73.7, 1.7) & \textbf{27.33\%} (8, 36.7, 7.3, 57.3) & \textbf{14.67\% (7.7, 4, 13.3, 33.7)} \\
        \hline
    \end{tabular}
    }
\end{table}

\begin{table}
    \centering
    \caption{V-Ratio and \textit{Eventual Verified-Ratio} (EVR) of Information Class-specific \vsum\ summaries (with $\kappa = 0$).}
    \label{tab:classWiseVer}
    \resizebox{\columnwidth}{!}{
    \begin{tabular}{|l|c|c|c|c|c|c|c|c|}
        \hline
        \multirow{2}{*}{\textbf{Event}} & \multicolumn{2}{c|}{\textbf{Victims}} & \multicolumn{2}{c|}{\textbf{Investigation}} & \multicolumn{2}{c|}{\textbf{Other Event Specific}} & \multicolumn{2}{c|}{\textbf{Opinions}} \\
        \cline{2-9}
        & \textbf{V-Ratio} & \textbf{EVR} & \textbf{V-Ratio} & \textbf{EVR} & \textbf{V-Ratio} & \textbf{EVR} & \textbf{V-Ratio} & \textbf{EVR} \\
        \hline
        
        Charliehebdo & 1.000 & 1.000 & \textbf{0.522} & \textbf{0.778} & 0.800 & 0.867 & 1.000 & 1.000 \\
        \hline
        
        Germanwings & 1.000 & 1.000 & \textbf{0.500} & \textbf{1.000} & 0.632 & 0.842 & 0.714 & 0.929 \\
        \hline
        
        Ottawa & 0.800 & 0.800 & \textbf{0.444} & \textbf{0.611} & 0.684 & 0.947 & 0.800 & 0.867 \\
        \hline
        
        Sydney & 0.933 & 0.933 & \textbf{0.571} & \textbf{0.778} & 0.700 & 0.800 & 0.813 & 0.813 \\
        \hline
        \hline
        
        \textbf{Overall} & 0.933 & 0.933 & \textbf{0.509} & \textbf{0.792} & 0.704 & 0.864 & 0.832 & 0.902 \\
        \hline
    \end{tabular}
    }
    \vspace{-1em}
\end{table}

\subsection{Case Study - Paris Terror Attack (Oct'20)}
\label{subsec:caseStudy}
In order to evaluate the practical utility of our proposed system, we present a case study on a recent terror attack in the northwest suburb of Paris where a French teacher, who had shown his students cartoons of the Prophet Muhammad, was decapitated outside his school on $16^{th}$ October 2020. The suspect was shot dead by the Police minutes later\footnote{https://en.wikipedia.org/wiki/Murder\_of\_Samuel\_Paty}. We scraped all the tweet threads available within three days from the date of the event, and used \vsum \hspace{1pt} (trained on all four datasets, $\kappa=0$) to generate a summary of length 250 words. {\bf 80\%} of the summary tweets were found (manually) to contain verifiable content, with the top 3 tweets, all conveying diverse information, along with their \textit{relevance scores} shown below.

\begin{tcolorbox}[title=Paris Terror Attack (October 2020)]
\begin{enumerate}[label=\textbf{T\arabic*:}, leftmargin=*]
    \item \textsf{The Muslim who beheaded a teacher in a street in France waited outside the school and asked pupils to identify his target, anti-terrorism officials say.} -- \textbf{0.91}
    \item \textsf{In an horrific attack outside Paris a teacher having taught a class about 'Freedom of Expression' beheaded by an 18yr old student, a Chechen, born in Moscow who filmed his attack before being shot dead by police at the scene! 9 people arrested include his family members! $<$link$>$} -- \textbf{0.86}
    \item \textsf{The teacher brutally murdered for doing his job in France yesterday has been named as 47-year-old Samuel Paty. $<$link$>$} -- \textbf{0.81}
\end{enumerate}
\end{tcolorbox}

%% file: conclusion.tex
\section{Conclusion }
\label{sec:conclusion}
We present \vsum, the first end-to-end solution for generating trustworthy summaries from large volumes of crisis-related tweets. \vsum\ consists of two modules, a tweet summarizer and a rumour detector. Different from existing disaster-specific summarizers, we model tweet summarization as a supervised classification task. Also, in an attempt to address the gap between the vast body of literature on document summarization and comparatively limited research on social media summarization, we propose a novel way to modify the architecture of an existing extractive document summarization technique for summarizing tweets. For the auxiliary task of \textit{rumour detection}, we achieve new SOTA results on the PHEME dataset, by jointly learning stances of replies towards the source tweet, and accumulating rumour-specific signals from the entire information cascade. Finally, we achieve new SOTA results for the primary task of \textit{trustworthy summarization} by jointly training the two module in a hierarchical multi-task setup to exploit their interaction. While such an end-to-end thorough study lays the foundation, deploying our proposed solution to provide trustworthy updates in times of crisis would be one of our immediate future endeavours.
